# A New Algorithm for Data Compression Optimization

I Made Agus Dwi Suarjaya
Information Technology Department
Udayana University
Bali, Indonesia

*Abstract*— People tend to store a lot of files inside theirs storage. When the storage nears it limit, they then try to reduce those files size to minimum by using data compression software. In this paper we propose a new algorithm for data compression, called j-bit encoding (JBE). This algorithm will manipulates each bit of data inside file to minimize the size without losing any data after decoding which is classified to lossless compression. This basic algorithm is intended to be combining with other data compression algorithms to optimize the compression ratio. The performance of this algorithm is measured by comparing combination of different data compression algorithms.

*Keywords- algorithms; data compression; j-bit encoding; JBE; lossless.*

## I. INTRODUCTION

Data compression is a way to reduce storage cost by eliminating redundancies that happen in most files. There are two types of compression, lossy and lossless. Lossy compression reduced file size by eliminating some unneeded data that won't be recognize by human after decoding, this often used by video and audio compression. Lossless compression on the other hand, manipulates each bit of data inside file to minimize the size without losing any data after decoding. This is important because if file lost even a single bit after decoding, that mean the file is corrupted.

Data compression can also be used for in-network processing technique in order to save energy because it reduces the amount of data in order to reduce data transmitted and/or decreases transfer time because the size of data is reduced [1].

There are some well-known data compression algorithms. In this paper we will take a look on various data compression algorithms that can be use in combination with our proposed algorithms. Those algorithms can be classified into transformation and compression algorithms. Transformation algorithm does not compress data but rearrange or change data to optimize input for the next sequence of transformation or compression algorithm.

Most compression methods are physical and logical. They are physical because look only at the bits in the input stream and ignore the meaning of the contents in the input. Such a method translates one bit stream into another, shorter, one. The only way to understand and decode of the output stream is by knowing how it was encoded. They are logical because look only at individual contents in the source stream and replace common contents with short codes. Logical compression method is useful and effective (achieve best compression ratio) on certain types of data [2].

## II. RELATED ALGORITHMS

### A. Run-length encoding

Run-length encoding (RLE) is one of basic technique for data compression. The idea behind this approach is this: If a data item *d* occurs *n* consecutive times in the input stream, replace the n occurrences with the single pair *nd* [2].

RLE is mainly used to compress runs of the same byte [3]. This approach is useful when repetition often occurs inside data. That is why RLE is one good choice to compress a bitmap image especially the low bit one, example 8 bit bitmap image.

### B. Burrows-wheeler transform

Burrows-wheeler transform (BWT) works in block mode while others mostly work in streaming mode. This algorithm classified into transformation algorithm because the main idea is to rearrange (by adding and sorting) and concentrate symbols. These concentrated symbols then can be used as input for another algorithm to achieve good compression ratios.

Since the BWT operates on data in memory, you may encounter files too big to process in one fell swoop. In these cases, the file must be split up and processed a block at a time [3]. To speed up the sorting process, it is possible to do parallel sorting or using larger block of input if more memory available.

### C. Move to front transform

Move to front transform (MTF) is another basic technique for data compression. MTF is a transformation algorithm which does not compress data but can help to reduce redundancy sometimes [5]. The main idea is to move to front the symbols that mostly occur, so those symbols will have smaller output number.

This technique is intended to be used as optimization for other algorithm likes Burrows-wheeler transform.

### D. Arithmetic coding

Arithmetic coding (ARI) is using statistical method to compress data. The method starts with a certain interval, it reads the input file symbol by symbol, and uses the probability of each symbol to narrow the interval. Specifying a narrower interval requires more bits, so the number constructed by the algorithm grows continuously. To achieve compression, the algorithm is designed such that a high-probability symbol narrows the interval less than a low-probability symbol, with the result that high-probability symbols contribute fewer bits to the output [2].

Arithmetic coding, is entropy coder widely used, the only problem is its speed, but compression tends to be better than Huffman (other statistical method algorithm) can achieve [6]. This technique is useful for final sequence of data compression combination algorithm and gives the most for compression ratio.

### III. PROPOSED ALGORITHM

J-bit encoding (JBE) works by manipulate bits of data to reduce the size and optimize input for other algorithm. The main idea of this algorithm is to split the input data into two data where the first data will contain original nonzero byte and the second data will contain bit value explaining position of nonzero and zero bytes. Both data then can be compress separately with other data compression algorithm to achieve maximum compression ratio. Step-by-step of the compression process can be describe as below:

1. Read input per byte, can be all types of file.
2. Determine read byte as nonzero or zero byte.
3. Write nonzero byte into data I and write bit '1' into temporary byte data, or only write bit '0' into temporary byte data for zero input byte.
4. Repeat step 1-3 until temporary byte data filled with 8 bits of data.
5. If temporary byte data filled with 8 bit then write the byte value of temporary byte data into data II.
6. Clear temporary byte data.
7. Repeat step 1-6 until end of file is reach.
8. Write combined output data
    a) Write original input length.
    b) Write data I.
    c) Write data II.
9. If followed by another compression algorithm, data I and data II can be compress separately before combined (optional).

Figure 1 shows visual step-by-step compression process for J-bit encoding. Inserted original input length into the beginning of the output will be used as information for data I and data II size. As for step-by-step of the decompression process can be describe below:

1. Read original input length.
2. If was compressed separately, decompress data I and data II (optional).
3. Read data II per bit.
4. Determine whether read bit is '0' or '1'.
5. Write to output, if read bit is '1' then read and write data I to output, if read bit is '0' then write zero byte to output.
6. Repeat step 2-5 until original input length is reach.

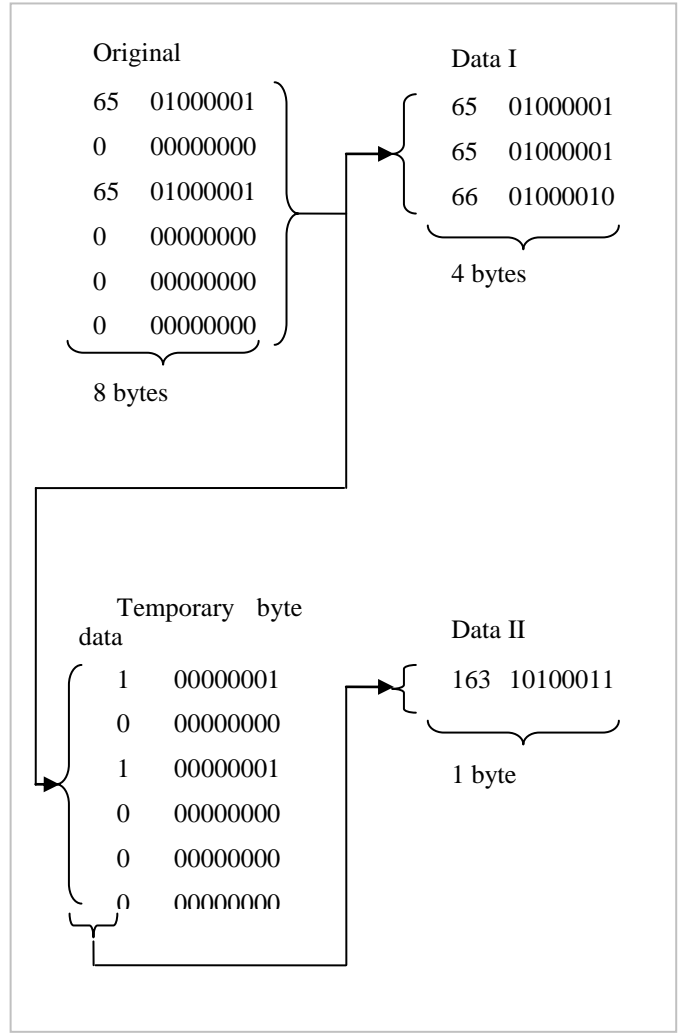

Figure 1. J-bit Encoding process

### IV. COMBINATION COMPARISON

Five combinations of data compression algorithm are used to find out which combination with the best compression ratio. The combinations are:

1. RLE+ARI.
2. BWT+MTF+ARI.
3. BWT+RLE+ARI.
4. RLE+BWT+MTF+RLE+ARI (as used in [3]).
5. RLE+BWT+MTF+**JBE**+ARI.

Those combinations are tested with 5 types of files. Each type consists of 50 samples. Each sample has different size to show real file system condition. All samples are uncompressed, this include raw bitmap images and raw audio without lossy compression. Average compression ratio for each type of file is used. Samples for the experiment are show in table 1.

TABLE I. SAMPLES FOR COMBINATION INPUT

| No | Name | Qty | Type | Spec. |
|---|---|---|---|---|
| 1 | Image | 50 | Bitmap Image | Raw 8 bit |
| 2 | Image | 50 | Bitmap Image | Raw 24 bit |
| 3 | Text | 50 | Text Document | |
| 4 | Binary | 50 | Executable, library | |
| 5 | Audio | 50 | Wave Audio | Raw |

## V. RESULT

Figure 2 shows that 8-bit bitmap images are compressed with good compression ratio by algorithms that combined with J-bit encoding.

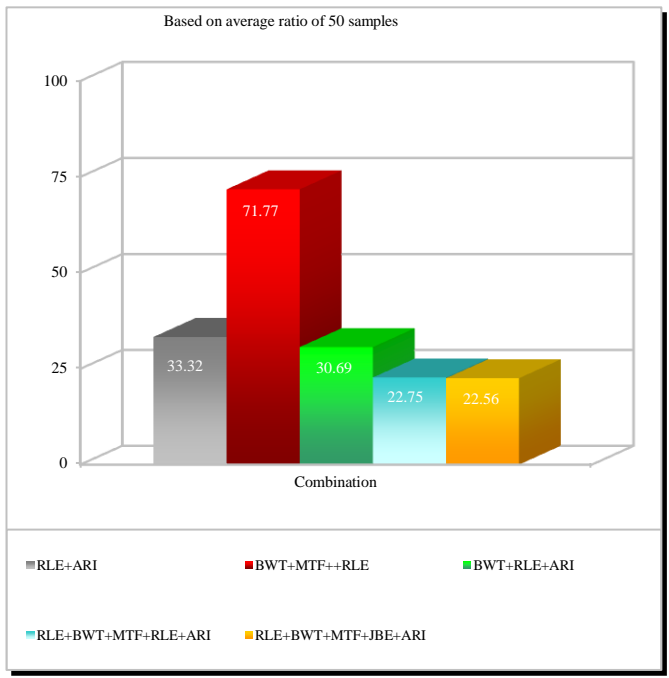

Figure 2. Ratio comparison for 8-bit bitmap image

Figure 3 shows that 24-bit bitmap images are compressed with better compression ratio by algorithms that combined with J-bit encoding. A 24 bit bitmap image has more complex data than 8 bit since it is store more color. Lossy compression for image would be more appropriate for 24 bit bitmap image to achieve best compression ratio, even thought that will decrease quality of the original image.

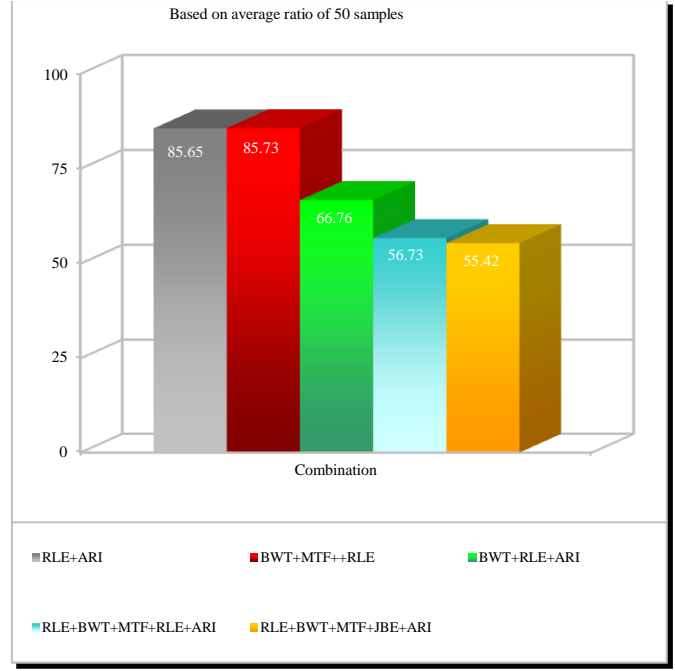

Figure 3. Ratio comparison for 24-bit bitmap image

Figure 4 shows that text files are compressed with better compression ratio by algorithms that combined with J-bit encoding.

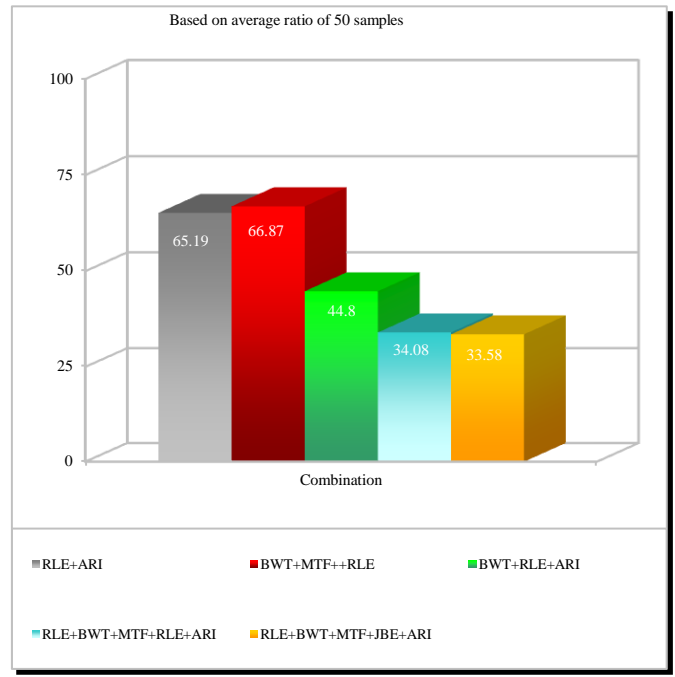

Figure 4. Ratio comparison for text

Figure 5 show that binary files are compressed with better compression ratio by algorithms that combined with J-bit encoding.

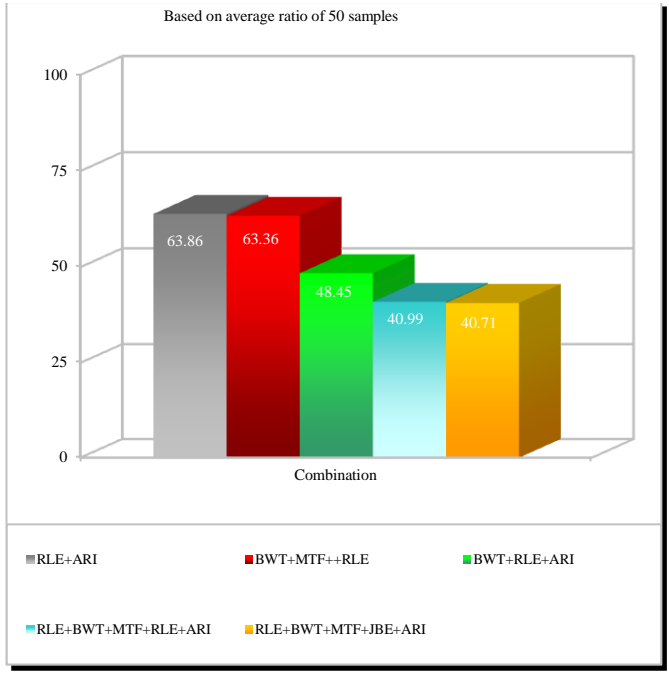

Figure 5. Ratio comparison for binary

Figure 6 shows that wave audio files are compressed with better compression ratio by algorithms that combined with J-bit encoding.

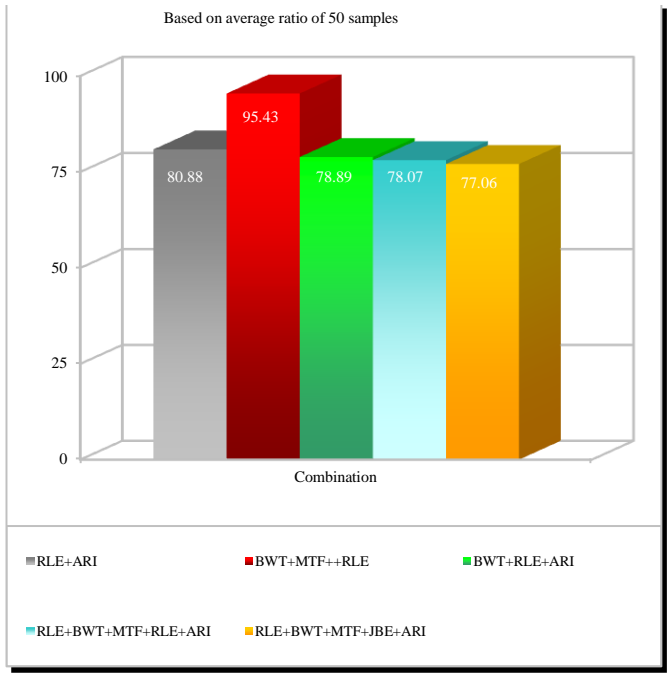

Figure 6. Ratio comparison for wave

## VI. CONCLUSION

This paper proposes and confirms a data compression algorithm that can be used to optimize other algorithm. An experiment by using 5 types of files with 50 different sizes for each type was conducted, 5 combination algorithms has been tested and compared. This algorithm gives better compression ratio when inserted between move to front transform (MTF) and arithmetic coding (ARI).

Because some files consist of hybrid contents (text, audio, video, binary in one file just like document file), the ability to recognize contents regardless the file type, split it then compresses it separately with appropriate algorithm to the contents is potential for further research in the future to achieve better compression ratio.

### REFERENCES

[1] Capo-chichi, E. P., Guyennet, H. and Friedt, J. K-RLE a New Data Compression Algorithm for Wireless Sensor Network. In Proceedings of the 2009 Third International Conference on Sensor Technologies and Applications.

[2] Salomon, D. 2004. Data Compression the Complete References Third Edition. Springer-Verlag New York, Inc.

[3] Nelson, M. 1996. Data compression with Burrows-Wheeler Transform. Dr. Dobb's Journal.

[4] Campos, A. S. E. Run Length Encoding. Available: http://www.arturocampos.com/ac_rle.html (last accessed July 2012).

[5] Campos, A. S. E. Move to Front. Available: http://www.arturocampos.com/ac_mtf.html (last accessed July 2012).

[6] Campos, A. S. E. Basic arithmetic coding. Available: http://www.arturocampos.com/ac_arithmetic.html (last accessed July 2012).

AUTHORS PROFILE

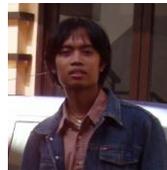

**I Made Agus Dwi Suarjaya** received his Bachelors degree in Computer System and Information Science in 2007 from Udayana University and Masters degree in Information Technology in 2009 from Gadjah Mada University. He served as a full-time lecturer at Faculty of Engineering, Information Technology Department in Udayana University. His research interest include software engineering, networking, security, computing, artificial intelligent, operating system and multimedia.